\documentclass[oneside,a4paper,reqno]{amsart}
\usepackage{amssymb}
\newtheorem{lemma}{Lemma}
\newtheorem{theorem}{Theorem}
\theoremstyle{remark}
\newtheorem*{remark}{Remark}
\begin{document}

\begin{titlepage}
\begin{center}
\bfseries  	THE BELL-KOCHEN-SPECKER THEOREM 
\end{center}
\vspace{1 cm}
\begin{center} D M APPLEBY
\end{center}
\begin{center} Department of Physics, Queen Mary
University of London,  Mile End Rd, London E1 4NS,
UK
 \end{center}
\vspace{0.5 cm}
\begin{center}
  (E-mail:  D.M.Appleby@qmul.ac.uk)
\end{center}
\vspace{0.75 cm}
\vspace{1.25 cm}
\begin{center}
\textbf{Abstract}\\
\vspace{0.35 cm}
\parbox{12 cm }{ 
  Meyer, Kent and Clifton (MKC) claim to have nullified
 the Bell-Kochen-Specker (Bell-KS) theorem.  It is true that they 
 invalidate KS's account of the theorem's physical implications.
 However, they do not invalidate Bell's point, that 
  quantum mechanics is inconsistent with the classical assumption, that
  a measurement tells us  about a property previously possessed
by the system. This failure of classical ideas about measurement is, perhaps,
the single most important implication of quantum mechanics.
  In a conventional colouring
there are some remaining patches of white.  MKC fill in these patches, but
only at the price of introducing patches where the colouring becomes
``pathologically'' discontinuous. The discontinuities mean that the colours in
these patches are empirically unknowable.  We prove a general theorem which
shows that their extent is at least as great as the patches of white in a
conventional approach. The theorem applies, not only to the MKC colourings,
but also to any other such attempt to circumvent the Bell-KS theorem
(Pitowsky's colourings, for example).  We go on to discuss the 
implications.    MKC do not
nullify the Bell-KS theorem.  They do, however, show that we did not,
hitherto, properly understand  the theorem.  For that reason their results
(and  Pitowsky's earlier results)  are of major importance.
   }
\end{center}
\end{titlepage}
\section{Introduction}
\label{sec:intro}
In ordinary language the word ``measurement''
\begin{quote}
very strongly suggests the ascertaining of some pre-existing property of some
thing, any instrument involved playing a purely passive role
\end{quote}
(in the words of Bell~\cite{Bell3}, p.166).  The Bell-Kochen-Specker
(Bell-KS) theorem~\cite{Bell1,Koch}  shows that
quantum mechanics is inconsistent with that
 natural idea.  

Or so Bell thought.  His conclusion has, however, been
challenged:  first by Pitowsky~\cite{PitowskyA,PitowskyB}, and then, using a
different set-theoretic argument, by Meyer~\cite{Meyer}, Kent~\cite{Kent1} and
Clifton and Kent~\cite{Kent2} (MKC).  MKC's argument was inspired by the
previous work of Hales and Strauss~\cite{Hales} and Godsil and
Zaks~\cite{Godsil}.  It has
attracted much
comment~\cite{Havli,Zeil1,Lars,me1,Cabel2,Breuer,MerminB,me2}.

MKC claim to have ``nullified'' the Bell-KS theorem.  They mean by
this that the  theorem, though mathematically valid, is not
physically  significant.  Pitowsky  expresses himself  less forcefully.
  He
does  not  say, in so many words, that the Bell-KS theorem is entirely
without significance.   However, he clearly means to insinuate doubts.

Now there can be no question as to the importance of the results proved by
Pitowsky and MKC (PMKC).  They clearly have some major consequences.  However,
we will argue that these consequences are less catastrophic than MKC think.
They do not show that the Bell-KS theorem is without significance.  They only
show  that we need to reassess its significance. 

The question is complicated by the fact that the Bell-KS theorem was
proved twice over:  first by Bell~\cite{Bell1}, and then again by Kochen and
Specker~\cite{Koch}.  They expressed strikingly different  views as
to the theorem's physical significance.  So the first thing one needs to
ask is:   what, exactly, is it that the MKC models are supposed to
nullify?

In Section~\ref{sec:Significance} we will argue that PMKC
have indeed invalidated Kochen and Specker's account of the theorem's
significance.   We will also argue that Bell's account  contains a number of
serious misconceptions.  So it is  true that PMKC nullify \emph{some} of
what was previously seen as the theorem's significance.  But they do not
nullify it all.   In particular, they do not nullify Bell's main point, as
stated above.  It remains the case that quantum mechanics
is inconsistent with classical ideas about measurement.

The fact that  PMKC  cannot have  fully restored the
ordinary, or classical concept of measurement becomes obvious, as soon as one
reflects that their models are still hidden variables theories.   
 
A hidden variables theory is one in which the pre-existing values are,
for some reason, concealed.  But if the observables  could be
measured in the ordinary, classical sense, then the values would not be
concealed.  They would be  open-to-view, as in classical
physics. It follows that, in a hidden variables theory, there must
necessarily be \emph{some} breakdown of classical assumptions about
measurement.

In Bohmian mechanics the values are concealed because measurements are
typically contextual.  Instead of the instrument playing a purely passive
role, as it would classically, there is a complex interplay between
system and instrument.  The effect is to 
\emph{create} a  value, which did not exist before.  As Bell
puts it:
\begin{quote}
  The result of a `spin measurement', for example, depends in a very
complicated way on the initial position $\lambda$ of the particle and on
the strength and geometry of the magnetic field.  Thus the result of the
measurement does not actually tell us about some property previously
possessed by the system, but about something  which has come into being
in the combination of system and apparatus. [Bell~\cite{Bell3}, p.~35]
\end{quote} 
(for a more detailed discussion of this point see Dewdney \emph{et
al}~\cite{DewdneyA} and Holland~\cite{Holland}).

PMKC have discovered a completely different mechanism for concealing
the 
values. 
  This is an important discovery.  It means, in
particular, that Bell's emphasis on the active role of the apparatus
needs revision.  Nevertheless, their models  still are hidden
variables theories.  So Bell's main point, that a measurement outcome
``does not actually tell us about some property previously possessed by
the system,'' still stands.

The variables are hidden in the PMKC models because their colourings are
violently, and even ``pathologically'' discontinuous (the relevance of
continuity in this regard is noted by Mermin~\cite{MerminB}).

Consider, for instance, the colouring described by Meyer~\cite{Meyer}.  This
function is discontinuous at every point in its domain of definition. It is,
in other words, as discontinuous as a function can possibly get. 

In a conventional colouring, such as a political map of the Earth, the paints
are  applied in broad strokes to well-behaved regions having regular
boundaries.  This means that it is almost always possible, using finite
precision measurements, to find out what country one is in.  It is true that
infinite precision would be needed if one was situated exactly on a boundary. 
However, there is zero probability of hitting such a point.

Meyer's colouring is not like that. 
In the Meyer colouring it is as if, before applying the paints,
 one first mixes
them, to the maximum extent possible, 
so that the colours become intermingled
at the molecular 
level.
It is (so to speak) a maximum entropy colouring.  This
makes the colours 
unobservable.

At least, the colours are not observable using finite precision measurements. 
MKC  say that finite precision nullifies
the KS~theorem.  But it would be nearer the truth if one said that finite
precision 
\emph{saves} the KS~theorem.

Suppose one tries to find out the value of some particular vector $\mathbf{n}
\in S^2_{\mathbb{Q}}$ (where $S^2_{\mathbb{Q}}$ is the  rational unit 2-sphere,
on which the Meyer colouring is defined) using a finite precision measurement.
The finite precision means that the measurement actually reveals the value of
some unknown vector $\mathbf{n}'$, with $|\mathbf{n}'-\mathbf{n}|\le \epsilon$
for some positive $\epsilon$.  If the colouring were continuous at
$\mathbf{n}$, and if
$\epsilon$ were sufficiently small, then the value of $\mathbf{n}'$ would be
the same as the value of $\mathbf{n}$.  However, the colouring is, in fact,
discontinuous.  This means that, no matter how small the  error
$\epsilon$, the measurement provides no more information about the value of
$\mathbf{n}$ than could be obtained by simply guessing a number at random
(assuming that $\epsilon$ is not actually $0$).  
The value  is therefore unobservable, or hidden.

A procedure which leaves the experimenter in complete ignorance of the
pre-existing values is clearly not a measurement in the classical sense.
 MKC focus on the point  that, in their models, a measurement does always
reveal the pre-existing value of
\emph{something}:  namely, the value of the vector $\mathbf{n}'$ representing
the true alignment of the instrument.  What they overlook is that the
experimenter does not know the true alignment of the instrument. 
This means that, although the experimenter learns a value, s/he has no idea
what it is a value \emph{of}. Consequently, the experimenter does not acquire
any actual knowledge.
 
A classical measurement is not \emph{simply} a procedure which reveals a
pre-existing value.  Rather, it is a procedure which ascertains a
pre-existing fact, of the form ``observable $A$
had value $x$''.    The specification of the observable $A$ is no less
essential than the specification of the value $x$.  The problem with the PMKC
models is that the observable $A$ is  not specified, so the experimenter 
only learns ``\ \dots\ had value $x$''.  This statement is completely 
uninformative.  It says no more than the statement ``observable $A$ had
\ \dots\ ''.  Indeed, it says no more than the completely empty statement
``\ \dots\ had\ \ \dots\ ''.

What emerges from this is that PMKC have  been asking the wrong question. 
The  important question is not:  ``How much of $S^2$ (the full unit
$2$-sphere) can be coloured \emph{at all}?''  But rather:  ``How much of
$S^2$ can be coloured in such a way that the colours are \emph{empirically
knowable}?''

In
Sections~\ref{sec:Preliminaries}--\ref{sec:Cabello} we address that
question.  We have seen that, in the case of Meyer's model, \emph{none} of
the colours are empirically knowable.  We need to consider whether another
model might improve on that.

So as to have a standard of comparison we begin, in
Section~\ref{sec:Preliminaries}, by defining the concept of a regular
KS-colouring.  Intuitively, this is a colouring where the paint is applied in
broad strokes, as in a political map of the Earth.   We show that a regular
KS-colouring must exclude a region having non-empty interior and subtending
solid angle 
$\ge 4\pi d_{\mathcal{R}}$.  Here 
$d_{\mathcal{R}}$ is a fixed positive number whose value is determined, once
and for all, by the principles of quantum mechanics.

We refer to functions of the same general kind as the PMKC colourings as
pseudo-KS-colourings of $S^2$.    In Section~\ref{sec:PseudoKS} we identify
two  conditions satisfied by every PMKC colouring (both the ones constructed
by Pitowsky and the ones constructed by MKC).  We argue that they would  also
have to be satisfied by any other pseudo-KS-colouring.

In
Sections~\ref{sec:Phenomenological}--\ref{sec:Discontinuity} we use these
conditions to analyze the discontinuities of an arbitrary pseudo-KS-colouring.
We show that $S^2$ splits into an open set $U$, which is \emph{regularly}
KS-colourable, and a closed set $D$ on which the discontinuities make the
colours empirically unknowable.  In the case of Meyer's colouring
$U$ is empty and $D=S^2$.  In the general case $D$ might be smaller.  However,
it cannot be
shrunk to nothing.  The fact that $U$ is regularly KS-colourable means that $D$
must always have non-empty interior, and subtend solid angle $\ge 4\pi
d_{\mathcal{R}}$.

In  Section~\ref{sec:NotNullified} we infer that the Bell-KS theorem is not
nullified.  A conventional colouring must exclude a region $D$ which is simply
not coloured at all.  PMKC have found ways to extend the colouring into $D$. 
However, they only do so at the price of making the valuation so extremely
discontinuous that the colours are empirically unknowable.  From the
point of view of a finite precision experimenter, who wants to ascertain the
pre-existing values, this is not an improvement.

The fact that the colours cannot all be empirically knowable is also shown
by Cabello~\cite{Cabel2}.  However, the relation between
Cabello's argument and ours may not be immediately apparent.
In Section~\ref{sec:Cabello} we elucidate the relationship. 

Finally, in   Section~\ref{sec:Significance} we assess the
implications of PMKC's discoveries, and our counter-argument.  
PMKC clearly  nullify \emph{some} of what used to be seen as the Bell-KS
theorem's significance.  We will argue that they completely invalidate what KS
say on the subject.  They also invalidate some of the things said by Bell.  In
addition we present some further criticisms of Bell, which are only indirectly
inspired by PMKC's argument.  In short, PMKC make us recognize that we did
not, in the past, fully understand what the  theorem is telling us.

However, none of this detracts from the theorem's importance.  The failure of
classical assumptions about measurement is arguably the single most
revolutionary feature of quantum mechanics.

\section{Regular KS-Colourings} 
\label{sec:Preliminaries} 
We begin by showing that there is certainly no question of nullifying the
Bell-KS theorem by means of the kind of well-behaved colouring one sees in a
political map of the Earth.  The results proved in this section will also
play an important role in our subsequent analysis of the PMKC models.

The Bell-KS~theorem states that there is no valuation $f\colon S^2 \to
\{0,1\}$ (where $S^2$ is the unit 2-sphere) such that
$f(-\mathbf{n})=f(\mathbf{n})$ for all
$\mathbf{n}$ and
\begin{equation}
f(\mathbf{n}_1) + f(\mathbf{n}_2) + f(\mathbf{n}_3) =2
\label{eq:KSCond1}
\end{equation}
for every triad (every  triplet of orthogonal unit vectors) $\mathbf{n}_1$,
$\mathbf{n}_2$,
$\mathbf{n}_3$.

The problem exposed by PMKC is that this statement, as it stands, is very weak.
It asserts that one cannot KS-colour  \emph{absolutely}  all of $S^2$.  But it
does not place any stronger constraint on the maximum size of a KS-colourable
set.  It consequently leaves PMKC free to argue that one can KS-colour 
\emph{effectively}  all of $S^2$, in some suitably defined sense of the
word ``effectively''.  

In this section we make a first step in the direction of strengthening the
theorem, by establishing constraints on the sizes 
of some special kinds of  KS-colourable set.

We will  have occasion to 
consider
\begin{enumerate}
\item the class $\mathcal{B}$ consisting of all
KS-colourable Borel sets (\emph{i.e.}\ all KS-colourable sets which are
measurable with respect to the usual rotationally invariant measure on
$S^2$).
\item the
class $\mathcal{C}$ consisting of all closed KS-colourable sets.
\item  the class $\mathcal{O}$ consisting of all open KS-colourable
sets.
\end{enumerate}
However, we are mainly interested in the class $\mathcal{R}\subseteq
\mathcal{O}$ consisting of all \emph{regularly} KS-colourable sets. 
Intuitively, a regular KS-colouring is one which exhibits the same kind of
``good'' behaviour one sees in a political map of the Earth.   Formally, a
KS-colouring 
$f\colon U \to \{0,1\}$ is regular if $U$ is open and $f$ is almost everywhere
continuous on $U$.

Before proceeding further we ought to remove a potential ambiguity.  We
consider  a function $f\colon K \to \{0,1\}$ defined on a  subset $K \subset
S^2$ to be a KS-colouring if and only if
\begin{enumerate}
\item $f(-\mathbf{n}) = f(\mathbf{n})$ whenever $-\mathbf{n}$ and
$\mathbf{n}$ both $\in K$.
\item $f(\mathbf{n}_1) + f(\mathbf{n}_2) + f(\mathbf{n}_3) =2$ for every 
triad $\mathbf{n}_1$, $\mathbf{n}_2$,
$\mathbf{n}_3 \in K$
\item  $f(\mathbf{n}_1) + f(\mathbf{n}_2) \ge 1$ for every orthogonal pair
$\mathbf{n}_1$,
$\mathbf{n}_2\in K$
\end{enumerate}
We do \emph{not} require that linear combinations of pairs of orthogonal
vectors both
$f$-evaluating to $1$ should also $f$-evaluate to $1$ (except, of course,
when that is implied by condition~$2$). The reason for not imposing this
additional requirement will appear  in 
Section~\ref{sec:PseudoKS}.

We begin by proving

\begin{lemma}
\label{lem:KSStrongerA}
If $K \in \mathcal{C}, \mathcal{O}$ or $\mathcal{R}$ then  the complement 
$S^2-K$ has non-empty interior.
\end{lemma}
\begin{remark}
In other words $K$ must exclude at least one
non-empty disk of the form $\{\mathbf{n} \in S^2\colon
\cos^{-1} \left(\mathbf{n}\cdot \mathbf{n}_0\right) < r\}$ with centre
$\mathbf{n}_0$ and radius $r>0$.  
\end{remark}
\begin{proof}
If $K \in \mathcal{C}$ the statement is immediate, since  $S^2 -K$ is then
open (being the complement of a closed set) and non-empty (in view of the
Bell-KS theorem).

If, on the other hand, $K\in \mathcal{O}$ or $\mathcal{R}$ we have to work a
little harder.

Let $\{\mathbf{n}_1,\mathbf{n}_2,\dots , \mathbf{n}_L\}$ be any finite 
KS-uncolourable set.  The fact that $\emph{K}$ is KS-colourable means
that at least one of these vectors must belong to $S^2-K$.  We may assume the
labelling is such that, for some $l\ge 1$, $\mathbf{n}_{i} \in S^2-K$ 
if $i \le l$ and $\mathbf{n}_{i} \in K$ if $i > l$.

If $l <L$, then the fact that $K$ is open means that we can choose $\delta >0$
such that $\{\mathbf{m} \in S^2 \colon \cos^{-1}(\mathbf{m} \cdot
\mathbf{n}_i)< 
\delta\}
\subseteq K$ for $i=(l+1), \dots L$.  If  $l = L$ then  we
choose
$\delta$ arbitrarily,
$=\pi/2$ say. 

If  
$\mathbf{n}_i$ is in the interior of $S^2-K$ for some $1\le i \le l$, 
then the interior of $S^2-K$ is non-empty, and the claim is proven.

Otherwise the vectors $\mathbf{n}_1, \dots , \mathbf{n}_l$ are all on the
boundary of $S^2-K$.   We can then  choose a
rotation, through an angle $<\delta$, which moves some 
of the vectors 
$\mathbf{n}_1, \dots , \mathbf{n}_l$ out of $S^2 - K$, without moving any of
the vectors  $\mathbf{n}_{l+1}, \dots, \mathbf{n}_{L}$ out of $K$.
After suitable re-labelling this gives us a new KS-uncolourable set 
$\{\mathbf{n}'_1,\mathbf{n}'_2,\dots , \mathbf{n}'_L\}$ with the
property that,  for some 
$1\le l' < l$,  
$\mathbf{n}'_{i} \in S^2-K$ if $i \le l'$ and $\mathbf{n}'_{i} \in K$ if $i >
l'$.

If it should still happen that none of the vectors
$\{\mathbf{n}'_1,\mathbf{n}'_2,\dots , \mathbf{n}'_{l'}\}$ is in the interior 
of $S^2-K$ we may repeat the procedure.  It is impossible to move all the
vectors out of $S^2-K$ so after sufficiently many iterations at least one of
the vectors must be in the interior of $S^2-K$.

Consequently, the interior of $S^2-K$ cannot be empty.
\end{proof}

{\allowdisplaybreaks
Our second result concerns the maximum area of a KS-colourable set.
Define
\begin{align}
d_{\mathcal{B}} 
& =1-\sup_{B\in \mathcal{B}}\bigl(\mu(B)\bigr)
\label{eq:dBdef}
\\
d_{\mathcal{C}} 
& =1-\sup_{C\in \mathcal{C}}\bigl(\mu(C)\bigr)
\label{eq:dCdef}
\\
d_{\mathcal{O}} 
& =1-\sup_{U\in \mathcal{O}}\bigl(\mu(U)\bigr)
\label{eq:dOdef}
\\
d_{\mathcal{R}} 
& =1-\sup_{U\in \mathcal{R}}\bigl(\mu(U)\bigr)
\label{eq:dRDef}
\end{align}
where $\mu$ is the usual rotationally invariant measure on $S^2$, normalized 
so that $\mu(S^2)=1$ (in other words, the solid angle scaled by $1/4\pi$).}
We will refer to these numbers as deficits.  They tell us the size of the
region excluded by a colouring of maximal extent.

In Appendix~\ref{sec:Lem2Proof} we prove
\begin{lemma}
\label{lem:KSStrongerB}
$d_{\mathcal{B}}$ is strictly $>0$.
\end{lemma}
We know that $d_{\mathcal{R}} \ge d_{\mathcal{O}} \ge d_{\mathcal{B}}$ and 
$d_{\mathcal{C}} \ge d_{\mathcal{B}}$ 
(because $\mathcal{R} \subseteq \mathcal{O}\subseteq \mathcal{B}$ and
$\mathcal{C}\subseteq \mathcal{B}$).
It can also be shown (see,
for example,  Halmos~\cite{Halmos}, Chapter~10) that
each Borel set
$B$ contains a sequence of closed subsets $C_n$ such that
$\mu(B) = \lim_{n\to \infty}\left(\mu(C_n)\right)$. Consequently
$d_{\mathcal{C}} \le d_{\mathcal{B}}$.  
Putting these facts together we deduce
\begin{equation}
d_{\mathcal{R}}\ge d_{\mathcal{O}} \ge d_{\mathcal{C}} = d_{\mathcal{B}} >0
\label{eq:deficits}
\end{equation}

It would be interesting to know whether the two inequalities at the left-hand
end of the chain are actually strict.  That is a question which requires
further investigation.

It would also be interesting to know the  values of the deficits.  It
is  easy\footnote{
  Simply assign  $0$ to the two  polar caps defined by  $|\tan
   \theta|
  <1$ and $1$ to the equatorial region defined by $|\tan
  \theta| >
  \sqrt{2}$ (where $\theta$ is the usual polar angle).  It is easily verified
  that this gives a regular KS-colouring covering a region having
  $\mu$-measure
  $=1-1/\sqrt{2} +
  1/\sqrt{3}$.
}
to
construct a regular KS-colouring which covers $87\%$ of $S^2$.  So we know
that
$d_{\mathcal{R}}$ (and therefore $d_{\mathcal{O}}$ and
$d_{\mathcal{B}}$) must be $\le 0.13$. In Appendix~\ref{sec:Lem2Proof} we show
that
$d_{\mathcal{B}}$ is bounded from below by an integral defined in terms of a
finite KS-uncolourable set. For the sets which have been described in the
literature this integral is rather small.  For instance, in the case of the
Conway-Kochen set~\cite{PeresBk,BubBk} the integral is $\lesssim
0.01$ (see Appendix~\ref{sec:Lem2Proof}).  So the possibility is not excluded
that 
$\sim 99\% $ of the total solid angle can be covered with a KS-colourable
Borel set (in terms of the Earth this would  correspond to colouring
everything except a region about the size of Australia). 

One should, however, note that the integral  is sensitive to the
angular separation of the rays in the corresponding KS-uncolourable set.  It
follows that, if one could find a set containing many fewer rays than the
Conway-Kochen set, this would be likely to give a subsantially larger lower
bound.  In any case, the integral is only a bound on $d_{\mathcal{B}}$, not
the actual value.  Lastly, it is possible that the deficits are larger in
higher dimensions.

The results we prove are enough to establish
that the Bell-KS theorem is not  nullified. 
However, it remains an open question, quite how close the theorem gets to
being nullified.  This too is a point that requires further investigation.

\section{Pseudo-KS-Colourings}
\label{sec:PseudoKS}
We refer to functions like the ones constructed
by PMKC as pseudo-KS-colourings of $S^2$.  In this section we identify two
conditions which the PMKC colourings all satisfy (both the ones constructed by
Pitowsky and the ones constructed by MKC).  We argue that they would also
have to be satisfied by any other pseudo-KS-colouring.

We aim to give a completely general
analysis, applying to any pseudo-KS-colouring.  But let us
begin by looking at the particular colourings constructed by PMKC.

The clearest and most succinct description of the Pitowsky colourings is in
Pitowsky~\cite{PitowskyA}.  Pitowsky~\cite{PitowskyB} contains important
additional material concerning the measure-theoretic aspects.  The interested
reader should also consult Pitowsky~\cite{PitowskyC,PitowskyD} (which concern the
Bell inequalities) and comments by Mermin and Macdonald~\cite{PitowskyComment}.

Pitowsky's approach is to define a function $f\colon S^2 \to \{0,1\}$
 on the whole of $S^2$.  This means  he has to relax the
requirement, that  $f$ should sum to $2$ on every triad.
Instead, he imposes the weaker requirement, that $f$ should sum to $2$ on
\emph{almost} every triad (in a sense of the word ``almost'' which is
not the standard measure-theoretic sense---see below).

Specifically, Pitowsky shows that there exist functions
$f\colon S^2 \to \{0,1\}$ such that 
\begin{enumerate}
\item
$f(-\mathbf{n})=f(\mathbf{n})$ for all $\mathbf{n}$
\item 
For all
$\mathbf{n}$, there are at most countably many orthogonal pairs
$\mathbf{m}, \mathbf{l} \in \mathbf{n}^\bot$  for which
\begin{equation}
f(\mathbf{n})+f(\mathbf{m}) + f(\mathbf{l}) \neq 2
\label{eq:KSCondPit}
\end{equation}
\end{enumerate}
 (where $\mathbf{n}^{\bot}$ is the orthogonal complement of
$\mathbf{n}$).

Let $T$ be the space of all triads, and let $N\subset T$ be the set of
``wrongly'' coloured triads.  A Pitowsky colouring has the property that
$p(N | \mathbf{n})=0$ for
all $\mathbf{n} \in S^2$ (where $p(N|\mathbf{n})$ is the conditional
probability of selecting a ``wrongly'' coloured triad, given that 
$\mathbf{n}$ is one of its elements).  Pitowsky infers that 
$p(N) =0$, so that almost all triads are
``correctly'' coloured.

It should be stressed  that there is a problem with this argument.  
Although  $p(N|\mathbf{n})$
is well-defined 
(in terms of the usual invariant
measure on the circle), there is a serious difficulty with the definition of 
$p(N)$. This is because $N$ is not a Borel set (as follows from a
variant of Lemma~\ref{lem:KSStrongerB} in the last section, applying to $T$
instead of $S^2$).
 Pitowsky  consequently has to rely on
``a strange concept of probability which violates the axiom of
additivity''~\cite{PitowskyD}.  For that reason his results have
 attracted  less attention than the subsequent work of 
MKC\footnote{
  Nevertheless, they are still worth considering:  not only for 
  reasons of completeness, but also because of their
  own intrinsic interest.  In particular, Pitowsky's attempt  
  to extract deep physical meaning from the \emph{arcana} of axiomatic
  set theory is, to our mind, intriguing. 

  Pitowsky's non-standard concept of probability 
  is certainly a source of serious difficulty. On the other hand,
  there is some force to his contention that,
    since $p(N|\mathbf{n})$ is well defined,
    it ought to be possible to make sense of $p(N)$ also. 
  Perhaps it is true 
  that
  one cannot make sense of $p(N)$ while remaining within the framework
  of the Kolmogorov axioms.  But that could be countered by asking
  whether there is any cogent physical reason which compels
  us to accept the Kolmogorov axioms (see 
  Pitowsky~\cite{PitowskyC,PitowskyD}).
}.

MKC's achievement was to find a way of obviating this difficulty.  Their
argument is formulated exclusively in terms of the standard theory of
probability, as formalized by the Kolmogorov axioms.  It therefore compels us
to take Pitowsky's suggestion, that the Bell-KS theorem may not have the
implications usually imputed to it, much more seriously.

Pitowsky assigns values to the whole of
$S^2$.  Meyer's~\cite{Meyer} key insight  is that we do not need to colour  the
whole of $S^2$ in order to account for the observations.  We might, for 
instance, be living in a world where
the only physically possible alignments are those specified by vectors
$\mathbf{n}
\in S_{\mathbb{Q}}^2$ (\emph{i.e.}\ unit vectors $\mathbf{n}$ whose
components are all rational).  

An experimenter may, indeed, set out with the
\emph{intention} of measuring in a direction $\mathbf{n}
\notin S_{\mathbb{Q}}^2$.  However, the finite precision of real laboratory
measurements means that we can never exclude the possibility that s/he
actually measures along a slightly different direction $\mathbf{n}'$ which
does $\in S_{\mathbb{Q}}^2$. The set
$S_{\mathbb{Q}}^2$ is KS-colourable.  Furthermore, the set of rational triads
is dense in the space of all real triads.  Meyer concludes that, since in this
model a measurement does always reveal the pre-existing value of the vector
which is
\emph{actually} measured, the Bell-KS theorem is nullified.  

Kent~\cite{Kent1}  subsequently extended Meyer's results to higher dimensional
spaces.  Clifton and Kent~\cite{Kent2} then showed that, in the case of finite
dimensional systems, models of the same general type can reproduce all the
statistical predictions of quantum mechanics, in so far as these are
verifiable by finite precision measurements (though it should be noted that
the Clifton-Kent models are still incomplete in that they do not specify the
dynamical behaviour of the 
system).

We want to give an argument which applies quite generally, not only to the
particular constructions of Pitowsky and MKC, but also to any other model
which might conceivably be thought to nullify the Bell-KS theorem.
 We therefore need to identify some minimal conditions which a
function
$f\colon K \to \{0,1\}$ must satisfy, if it is to count as a
pseudo-KS-colouring of $S^2$. 

 Our first such condition is the following:

\medskip
\parbox{4.6 in}{
\textbf{Condition $\mathbf{1}$.}
The domain $K$ is a dense subset of $S^2$.
}

\medskip \noindent
We consider this condition to be necessary because,  if it is not satisfied,
there is a non-empty disk which is not coloured at all.
It should be noted that we do not require that $K$ is countable   (as
in the MKC colourings).  In particular, the possibility is not excluded that
$K = S^2$ (as in the Pitowsky colourings).

We do not require  $f$ to be a KS-colouring of $K$.  Nor do we require it to
be Borel-measurable.  We do, however, require

\medskip
\parbox{4.6 in}{
\textbf{Condition $\mathbf{2}$.}
Suppose that $\mathbf{n}_1, \mathbf{n}_2, \mathbf{n}_3$ is a triad of
vectors $\in S^2$ (they need not $\in K$), and suppose that, for each $r$,
$U_r$ is an open neighbourhood of $\mathbf{n}_r$ with the property $K \cap
U_r \subseteq f^{-1} (\{a_r\})$ for  $a_r = 0$ or $1$.  Then
\begin{equation}
  a_1+a_2 +a_3 = 2
\nonumber
\end{equation}
Similarly, if  $\mathbf{n}_1, \mathbf{n}_2$ is an orthogonal pair of
vectors $\in S^2$ (not necessarily $\in K$),  and if, for each  $r$, 
$\mathbf{n}_r$ has an open neighbourhood 
$U_r$ such that $K \cap
U_r \subseteq f^{-1} (\{a_r\})$ for  $a_r = 0$ or $1$, then
\begin{equation}
  a_1+a_2  \ge 1
\nonumber
\end{equation}
}

\medskip  

\noindent
The reader may  verify that the colourings described by PMKC 
satisfy this condition.
We consider it to be necessary because, if it is not satisfied,
there exist finite precision measurements whose outcomes are guaranteed to
conflict with the predictions of quantum 
mechanics\footnote{
  This is the reason we adopted the less restrictive
  definition of a KS-colourable set in Section~\ref{sec:Preliminaries}.  
  It is not obvious that the same would be true of the version
  of condition~2 corresponding to the more restrictive definition.
}.

Let us stress that these are minimal conditions, which every
pseudo-KS-colouring must satisfy.  They are not intended to fully characterize
the concept.

\section{The Phenomenological Colouring}
\label{sec:Phenomenological}
The mathematical properties of a pseudo-KS-colouring are best visualized
using a three-coloured variant of the usual chromatic metaphor.  We will
call this the phenomenological colouring.  Intuitively, it describes what is
seen when the sphere is viewed through finite resolution eyes.  

Consider some fixed pseudo-KS-colouring $f\colon K\to \{0,1\}$.   Let
$K_0= f^{-1}(\{0\})$  be the set of points $f$-evaluating to $0$, and let
$K_1=f^{-1}(\{1\})$ be the set of points $f$-evaluating to $1$.

 We
begin by defining a true, or intrinsic colouring. We take points $\in K_0$
to be intrinsically blue, and points $\in K_1$ to be intrinsically red. 
Points $\notin K$ (if any) we take to be intrinsically white. 

Suppose, now, that one has an open region which entirely consists of points 
intrinsically blue or white.  We take it that such a region, seen through
 finite resolution eyes, would appear \emph{phenomenologically blue} 
(all of it,
including the points which are intrinsically white).  Similarly, we take it
that an open region which entirely consists of points intrinsically red or
white would appear \emph{phenomenologically red} 
(all of it, including the points
which are intrinsically white).  

Suppose, on the other hand, that each neighbourhood of $\mathbf{n}$  contains
at least one intrinsically blue point, and at least one intrinsically red
point. Then we take $\mathbf{n}$  to be \emph{phenomenologically black} 
(irrespective
of whether it is in fact intrinsically blue, red or 
white).

One can think of the phenomenological colours as arising through the mixing
of the true or intrinsic 
paints\footnote{
  In practice a mixture of blue and red paint gives purple, not black.
  We, however, are considering   paint that
  is ideally blue (so that it is a perfect absorber for wavelengths $\ge
  491\ \mathrm{nm}$) and paint that is ideally red (so that it is a perfect
  absorber for wavelengths $ \le 647\ \mathrm{nm})$. 
}. 
 However, this mixing analogy, though
helpful on an intuitive level, should not be taken too far.  For
instance,  a single intrinsically blue point surrounded by a region which is
otherwise intrinsically pure red counts, on our definition, 
as phenomenologically black (not red as, one might argue, it would actually
appear).

Our aim in this is not really to describe the actual visual appearance of a
surface which has been stippled with differently coloured microscopic dots. 
We simply want to present a convenient  way to picture the abstract
mathematical properties of a pseudo-KS-colouring.

\section{The Discontinuity Region}
\label{sec:Discontinuity}
We now use the phenomenological colouring to investigate the
discontinuities of $f$. 

Let us begin by defining the phenomenological colouring in more formal
terms.  Let
$\overline{K}_0$ (respectively
$\overline{K}_1$) be the closure of $K_0$ (respectively $K_1$) considered  as
a subset of $S^2$. Then $\overline{K}_0 \cup \overline{K}_1 = \overline{K} =
S^2$.  Now define
$U_0 = S^2 - \overline{K}_1$, $U_1 = S^2 - \overline{K}_0$ and $D =
\overline{K}_0 \cap \overline{K}_1$.  Then  $U_0$, $U_1$, $D$ partition $S^2$
into three pairwise disjoint subsets.  Furthermore, $U_0$, $U_1$ are open and
$D$ is closed.  

The phenomenological colouring $\tilde{f} \colon S^2 \to \{0,1,-1\}$ may now
be defined formally, by
\begin{equation}
\tilde{f}(\mathbf{n}) =
\begin{cases}
 \phantom{-} 0 \qquad \text{if $\mathbf{n} \in U_0$ } \\
\phantom{-} 1 \qquad \text{if $\mathbf{n} \in U_1$} \\
 - 1 \qquad \text{if $\mathbf{n} \in D$} 
\end{cases}
\end{equation}
Thus, $\mathbf{n}$ is phenomenologically blue, red or black depending on
whether it $\tilde{f}$-evaluates to $0$, $1$, or $-1$ respectively.

This topological gambit enables us to escape all the difficulties
arising from the fact that $f$ is not assumed to be Borel-measurable. 
$K_0$, $K_1$ may or may not be Borel sets.   However,  $U_0$, $U_1$
(being open) and  $D$ (being closed) are guaranteed to be Borel sets. 
Consequently, $\tilde{f}$ is guaranteed to be Borel-measurable.

Let $U = U_0 \cup U_1$.  Then the true colouring $f$ is continuous at every
point of $K\cap U$, and discontinuous at every point of $K \cap D$.  We will
therefore refer to $U$ as the continuity region, and to $D$ as the
discontinuity region.

The significance  of these sets is that they describe the extent to which  the
true colours are empirically knowable.

Suppose $\mathbf{k} \in K\cap U$.  Then every vector $\in K$ sufficiently
close to $\mathbf{k}$ has the same intrinsic colour as $\mathbf{k}$. 
Consequently, a sufficiently accurate finite precision measurement in the
approximate direction of $\mathbf{k}$ is guaranteed to reveal that colour.

Suppose, on the other hand, that $\mathbf{k} \in K\cap D$.  Then each
neighbourhood of
$\mathbf{k}$, no matter how small, contains infinitely many other vectors $\in
K$  having the opposite intrinsic colour.  This means that
the true colour of $\mathbf{k}$ cannot be reliably ascertained by finite
precision measurement in the direction $\mathbf{k}$.

It is obvious that $D$ cannot be empty ($S^2$ is connected so it cannot be
the disjoint union of two non-empty open sets $U_0$ and $U_1$).  That would
not be a serious problem if $D$ was in some sense negligible (even a
political map of the Earth is discontinuous on the boundary lines).  However,
$D$ is, in fact,  non-negligible, as the following theorem shows.
\begin{theorem}
\label{thm:phenomenological}
Let $f\colon K \to \{0,1\} $ be a pseudo-KS-colouring of $S^2$, with
continuity  region $U$, and discontinuity region $D$.  Let $\tilde{f}_0\colon
U \to \{0,1\}$ be the restriction of the phenomenological colouring to $U$. 
Then
\begin{enumerate}
\item $\tilde{f}_0$ is a regular KS-colouring of $U$.
\item $D$ has non-empty interior.
\item $\mu (D) \ge d_{\mathcal{R}}$.
\end{enumerate}
\end{theorem}
\begin{remark}
$d_{\mathrm{R}}$ is the deficit defined in Eq.~(\ref{eq:dRDef}).
\end{remark}
\begin{proof}  
We only need prove the first of these statements.   The other two statements 
will then be  immediate consequences  of results proved in
Section~\ref{sec:Preliminaries} .

We note, first of all, that $\tilde{f}_0$ is continuous (because the sets 
$\tilde{f}_{0}^{-1} (\{0\})=U_0$ and $\tilde{f}_{0}^{-1} (\{1\})=U_1$ 
are both open).

Now let $\mathbf{n}_1, \mathbf{n}_2, \mathbf{n}_3$ be any triad of
vectors $\in U$.  Since $\tilde{f}_{0}$ is continuous we can choose, for each
$r$, an open neighbourhood $V_r$ of $\mathbf{n}_r$ such that $V_r \subseteq
U$ and 
$\tilde{f}_{0}(\mathbf{m}) = \tilde{f}_{0}(\mathbf{n}_r)$ for all $\mathbf{m} \in
V_r$.  Consequently $f(\mathbf{m}) = \tilde{f}_{0} (\mathbf{n}_r)$ for all
$\mathbf{m} \in K\cap V_r$.  It now follows from condition~2 in
Section~\ref{sec:PseudoKS} that
\begin{equation}
 \tilde{f}_{0} (\mathbf{n}_1) + \tilde{f}_{0} (\mathbf{n}_2) + 
\tilde{f}_{0} (\mathbf{n}_3) = 2
\end{equation}
In the same way we can show 
\begin{equation}
 \tilde{f}_{0} (\mathbf{n}_1) + \tilde{f}_{0} (\mathbf{n}_2) \ge  1
\end{equation}
for any pair of orthogonal vectors $\mathbf{n}_1, \mathbf{n}_2 \in U$.

This shows that $\tilde{f}_{0}$ is a KS-colouring.  The regularity is a
consequence of the fact that $\tilde{f}_0$ is continuous.
\end{proof}
Theorem~\ref{thm:phenomenological} is the central result of this paper. It
shows that, in so far as the aim is to maximize the area on which the true
colours are \emph{empirically knowable}, pseudo-KS-colourings 
do no better than
regular ones.

In the remainder of this section we examine the internal structure of 
$D$.  It should already be apparent  that the
discontinuities might fairly be described as ``pathological''.  However, in
the general case the situation is  more complicated than it is with
Meyer's colouring.

Let $D_{\mathrm{i}}$ be the interior of $D$.  In
$D_{\mathrm{i}}$ the valuation shows exactly the same kind of 
``pathologically'' discontinuous behaviour  as the Meyer
colouring.  Each $\mathbf{n}\in D_{\mathrm{i}}$ is surrounded by a disk which
is pure phenomenological black.  This means that the intrinsic blues and reds
are completely intermixed, everywhere, at the molecular level so
to speak.  In the case of the Meyer colouring $D_{\mathrm{i}}=S^2$.  In the
general case it may be smaller.  But it must always happen that
$\mu(D_{\mathrm{i}}) >0$.

If the boundary of $D$ is sufficiently well-behaved---if, for example, it
consists of a finite set of closed $C^{1}$ curves---then
$\mu(D_{\mathrm{i}}) =
\mu(D)$.  However, in the general case it might happen that
$\mu(D_{\mathrm{i}}) <
\mu(D)$ (it might, for example, happen that $D$ has an infinitely complex
lace-like structure, something like what one sees in the Mandelbrot
set~\cite{Devaney}).  We therefore need to examine the nature of the
discontinuities outside
$D_{\mathrm{i}}$. For that we can appeal to the following lemma.
\begin{lemma}
\label{lem:LebesgueVitali}
Let $f\colon K \to \{0,1\}$ be a pseudo-KS-colouring with discontinuity
region $D$.  Let $S(\mathbf{n},\epsilon) = \{\mathbf{m}\in S^2\colon
\cos^{-1} \left( \mathbf{m}\cdot \mathbf{n} \right) < \epsilon\}$ be the disk
with centre $\mathbf{n}$ and angular radius $\epsilon$.  Then the limit
\begin{equation}
 \lim_{\epsilon\to 0} \left(
\frac{\mu\bigl(S(\mathbf{n},\epsilon)\cap
D\bigr)}{\mu \bigl(S(\mathbf{n},\epsilon)\bigr)} 
\right)
\nonumber
\end{equation}
exists and $=1$ for almost all $\mathbf{n}\in D$.
\end{lemma}
\begin{proof}
See Appendix~\ref{ap:LebesgueVitali}.
\end{proof}
Let $D_{\mathrm{m}}$ be the subset of $D$ on which the limit exists and
$=1$.  It is easily seen that $D_{\mathrm{i}} \subseteq D_{\mathrm{m}}$.
We can think of $D_{\mathrm{m}}$ as the measure-theoretic  interior.  In some
ways it gives a better  idea of the interior, intuitively conceived, than
the set $D_{\mathrm{i}}$, defined topologically.

If $\mathbf{n} \in D_{\mathbf{m}}-D_{\mathbf{i}}$  the disk
$S(\mathbf{n},\epsilon)$ is not completely pure  black
for any positive
$\epsilon$ (otherwise $\mathbf{n}$ would $\in D_{\mathrm{i}}$).  However, if
$\epsilon$ is sufficiently small the disk is  \emph{nearly} pure black. 
Furthermore, the proportion of blackness comes arbitrarily close to $1$ as
$\epsilon\to 0$.  This means that the mixing of intrinsic blues and reds
in the vicinity of $\mathbf{n}$ is not entirely complete.  There are some
microscopic specks of phenomenological blue or red.  However, the mixing
is
\emph{nearly} complete---implying that the intrinsic valuation $f$ exhibits
\emph{nearly} the same degree of ``pathological'' discontinuity here that it
does on $D_{\mathrm{i}}$. 

The  intrinsic colours of  vectors $\in
K\cap(D_{\mathrm{m}}-D_{\mathrm{i}})$ cannot be ascertained by finite
precision measurement.  Let us note that the same is \emph{effectively} true
of the vectors in the specks of phenomenological blue or red buried deep in
the interstices of the region $D_{\mathrm{m}}-D_{\mathrm{i}}$.  Strictly
speaking these specks are contained in
$U$, not $D$.  However, as  one pushes in closer and closer to a vector
$\mathbf{n} \in D_{\mathrm{m}} - D_{\mathrm{i}}$ the specks become smaller
and smaller.  Consequently, the precision needed to ascertain the true
colours of the vectors in them grows without bound.  There will, for
instance, come a point when the precision needed is so large that it would
take a disk-pack the size of the observable universe to store the bit-string
specifying a single vector $\mathbf{k}$ to that degree of exactitude.  We may
 say that the true colours of such vectors  are
unobservable FAPP (unobservable ``for all practical purposes'').

We may thus visualize $D$ as consisting of three concentric layers:
\begin{enumerate}
\item An inner core $D_{\mathrm{i}}$ on which $f$  has the same
``pathologically'' discontinuous  character as Meyer's colouring.
\item An intermediate mantle $D_{\mathrm{m}}-D_{\mathrm{i}}$ on which $f$
has nearly the same ``pathologically'' discontinuous  character as Meyer's
colouring.
\item An outer crust $D-D_{\mathrm{m}}$ on which the discontinuities may be
comparatively mild.
\end{enumerate}
The crust $D-D_{\mathrm{m}}$ has
$\mu$-measure zero.  So we may
conclude that
$f$ is ``pathologically'' discontinuous over almost the whole of $D$.

Let us make one final point.  Even the mildest of discontinuities is enough
to frustrate the finite precision experimenter.  Consider two 
colourings (not KS-colourings) $g_{b}$ and $g_{r}$ each of
which assigns blue to everything north of the equator, and red to everything
south of it.  However, the equator itself is coloured blue by $g_{b}$
and red by
$g_{r}$.  Then
$g_{b}$ and
$g_{r}$ are empirically indistinguishable.

It is tempting to think of  a line discontinuity as somehow ``harmless''. 
This is true in the sense that  the line  has $\mu$-measure
zero (implying that there is zero probability of landing  on it).  But it
is still the case that  the line's true colour is
unobservable.

\section{The Bell-KS Theorem is not Nullified}
\label{sec:NotNullified}
In a regular KS-colouring there are some remaining patches of white, which are
simply not coloured at all.  A pseudo-KS-colouring replaces these patches of
white with patches of black, on which the colours are defined, but empirically
unknowable.  From the point of view of a finite precision experimenter, who
wants to ascertain the intrinsic colour of a specified point, this is not an
improvement.

Conventional models, such as the Bohm theory, and unconventional models, such
as the ones proposed by PMKC, make completely different statements about what
is going on ``behind the scenes''.   In a conventional model the
pre-existing values are concealed because the apparatus actively manufactures 
 new values.  In a PMKC model, by
contrast, the values
are concealed because the valuation is ``pathologically'' discontinuous.

The difference is important.  The PMKC models have major implications for the
way we understand the Bell-KS theorem, as we discuss in
Section~\ref{sec:Significance}.  However, these implications do not include
the statement, that the  theorem is nullified.  Bell's point, that a
quantum measurement ``does not actually tell us about some property previously
possessed by the system'' (Bell~\cite{Bell3}, p.35) remains intact.

\section{Cabello's Argument}
\label{sec:Cabello}
Cabello~\cite{Cabel2}, in an important paper, has given an argument which is
closely related to ours.    

Suppose an experimenter makes a finite precision measurement in the direction
$\mathbf{k}\in K$ with alignment uncertainty $\epsilon$.  Let
$p(\mathbf{k},\epsilon)$ be the probability that the measurement reveals the
true colour of $\mathbf{k}$.  Classically, one would assume 
\begin{equation}
\lim_{\epsilon\to 0} \bigl( p(\mathbf{k},\epsilon)\bigr) = 1
\label{eq:Cabello}
\end{equation}
Cabello, however, 
shows\footnote{
  We should note that Cabello  states the conclusion
  differently.  As he sees
 it his argument 
  shows that the MKC models ``lead to 
  experimentally testable predictions that are in contradiction with those
  of quantum mechanics''.  However, if one examines the last 
  paragraph in Section~IV of his paper, and  
  endnote~28 of his paper, it becomes clear that such  
  a  contradiction only arises if MKC's own assumptions are supplemented 
  with the additional assumption that ``successive tests with increasing 
  precision will give us the true colours with a higher probability''.

  We
  are unable to detect any mathematical error in Clifon and Kent's
  proof~\cite{Kent2}
  that their models do reproduce the experimentally testable
  predictions
  of quantum mechanics.  So
   we conclude that what Cabello actually does is to show, by
  \emph{reductio ad absurdum},
  that the MKC models are inconsistent with his
  additional assumption (and consequently with our Eq.~(\ref{eq:Cabello})).
  }
that the MKC models are inconsistent with that
natural assumption.  

Cabello's argument thus establishes, by a different route, the point we made
in Section~\ref{sec:Discontinuity}:  namely, that the colours in $K$ cannot
all be reliably ascertainable by finite precision measurement.

It would be interesting to see if Cabello's approach could be extended.  It
would, for instance, be interesting to investigate the size of the 
set\footnote{
  Let us note that our argument in 
  Sections~\ref{sec:Discontinuity}--\ref{sec:NotNullified}, though it
   suggests, does
  not logically imply  any statement regarding the size of this set.
  Nor does it tacitly depend on such a statement.

  Suppose  that, for some
  $\epsilon$ and some
  non-empty open $V\subseteq
   D$,
  $p(\mathbf{k}, \epsilon) =1$ for all $\mathbf{k} \in
  K\cap V$.  This would
  mean, in effect, that measurements to
  finite precision $\epsilon$ are really  infinite   precision measurements so
  far as the true colours are concerned.  But an experimenter could not
  exploit this fact to acquire information about the true colours because s/he
  could not  distinguish an instrument guaranteed to reveal the true 
  colour of $\mathbf{k}$ from another instrument guaranteed to reveal the
   true colour of a nearby vector $\mathbf{k}'$. 
}
on which $\lim_{\epsilon\to 0}
\bigl( p(\mathbf{k},\epsilon)\bigr)$ either fails to exist, or exists but is
$\ll 1$.

\section{The Physical Significance of the Bell-KS Theorem}
\label{sec:Significance}
The PMKC models show that the significance  of the Bell-KS theorem
is primarily epistemological:  it concerns the nature and extent of the
knowledge acquired by measurement.  In this section we examine how far that
proposition departs from  the views of Bell and KS.  We also
explain why, in our view, the theorem matters:  why it deserves its status as
one of the key foundational results of quantum mechanics.

For a long time  it was widely (though not universally) believed that quantum
mechanics is just plain inconsistent with the classical picture, of  particles
moving along sharply-defined, objective trajectories.  However, Bohm's 1952
rediscovery~\cite{Bohm} of de Broglie's pilot wave theory~\cite{Broglie} showed
that that is incorrect.  As Bell puts it:  ``in
$1952$ I saw the impossible done'' (Bell~\cite{Bell3}, p.160).

The Bell-KS theorem was conceived in response to that event.  Bell and KS,
in their different ways, were both trying to establish that a hidden variables
theory does not really mark the restoration of classical physics.  
However, their attitudes  could hardly have
been more divergent.  

Let us 
begin by examining KS's ``take'' on the theorem.
KS~\cite{Koch}
 describe the theorem as ``a proof of the non-existence of hidden
variables''.  This is unfortunate, for it is rather obviously no such
thing.  However, one finds on closer inspection that they do not really
mean that.  The intuition which drives their work is the perception that
there is a kind of logico-mathematical symmetry  to quantum mechanics, which
a hidden variables theory must violate.  They try to capture that intuition
by means of a formal criterion.  Specifically, they maintain that  a
``successful'' hidden variables theory must assign, to  each quantum
observable
$\hat{A}$, a real function $f_{\hat{A}}$ defined on a phase space $\Omega$
with the property 
\begin{equation}
f_{g(\hat{A})} = g \circ f_{\hat{A}}
\label{eq:KSfuncCond}
\end{equation}
for each Borel function $g$.  They use the Bell-KS theorem to infer that no
such assignment is possible and, consequently, that   
``successful'' hidden variables theories do not exist.

We will not say much more about this  because it is an
implication that PMKC clearly do invalidate. It is true that  functions
satisfying Eq.~(\ref{eq:KSfuncCond}) cannot be assigned to every  quantum
observable.  However, Clifton and Kent~\cite{Kent2} demonstrate, by explicit
construction, that such functions can be assigned to all the observables in a
dense subset.   KS  concede in advance that this is enough to
invalidate their argument because they accept
(Kochen and
Specker~\cite{Koch}, p.70)
\begin{quote}
that in fact it is not physically meaningful to assume that there are a
continuum number of quantum mechanical propositions 
\end{quote}

Before moving on let us say that, although KS's specific proposal has been
shown not to work, there might  be some substance to their underlying
intuition.  It certainly seems to this writer (on an intuitive level) that
there is a kind of symmetry to quantum mechanics, which hidden variables
 spoil.  It might  be worth trying to find a more satisfactory way to capture
that intuition formally.

Bell approaches  the problem from a completely different angle.  KS are
interested in questions of abstract logico-mathematical structure.  Bell, by
contrast, is motivated by  a strong philosophical objection to the
Copenhagen Interpretation. He particularly objects to the fact that the
Copenhagen Interpretation accords primacy to a concept, ``observation'',
which, besides being subjective, is not even sharply defined (see, for
example, Bell~\cite{Bell3}, p.~174).  It is probably fair to say  that he does
not regard the de Broglie-Bohm  theory   as a satisfactory
solution to the interpretation problem.  But he certainly sees it as an
improvement on the Copenhagen Interpretation.  This means that, where KS are
 looking for  reasons to rule out the hidden variables idea, Bell is
looking for clues which may guide us to a fully satisfactory, fully objective
interpretation of quantum mechanics.

As Bell sees it the theorem  shows
(Bell~\cite{Bell3}, pp.~8--9 and 164--6) that
measurement outcomes must depend, not only on the observable measured and the
hidden state of the system, but also on the complete experimental set-up (or
measurement context).  However, that does not (he thinks)  represent any kind
of objection to the hidden variables concept.  It simply means that hidden
variables theories are not classical theories, and so cannot be expected to
obey the classical rules.  Specifically (\emph{ibid.},
pp.~2, 9, 35, 165, 166) he sees  contextuality as the  manifestation, in hidden
variables terms, of  Bohr's~\cite{Bohr2}
point  concerning
\begin{quote}
  the impossibility of any sharp distinction between the behaviour of atomic
objects and the interaction with the measuring instruments which serve to
define the conditions under which the phenomena appear.
\end{quote}

Bell's anxiety that the theorem should not be seen as an impossibility proof
is so great that it seriously unbalances his exposition.  As
Mermin~\cite{MerminA} notes he seems, in places, almost to suggest that
the  theorem is ``silly''.  However, the following  
passage
shows
that he does not really think it ``silly'' 
(Bell~\cite{Bell3}
p.~166)\footnote{
  Bell does not explicitly mention 
  the Bell-KS theorem in this passage.   For some reason
  he seeks to minimize his own contribution throughout the
  paper~\cite{Bell4} from which it
  is taken. For instance, two pages earlier
  (Bell~\cite{Bell3}, p.~164)  he introduces his contextuality theorem as:
   ``\dots the Gleason-Jauch proof.  I was told of it by
   J.M.~Jauch
  in $1963$.  Not all of the powerful mathematical theorem of Gleason
  is required, but only a corollary  which is easily proved by itself. (The
  idea was later rediscovered by Kochen and Specker; see also
  Belinfante and Fine and Teller)''---as though he
  himself had nothing  to
  do
  with it. He gives  a long  list of names---Gleason, Jauch, Kochen,
  Specker, Belinfante, Fine, Teller---but omits to mention his 
  own (though it will be found that one of the numbered citations  is to
  Bell~\cite{Bell1}).  Quite why
   Bell should want to pass his theorem off,
  first as
  something Jauch told him, and then as something we learned from Bohr, is not
  entirely clear. Most probably it is connected with his desire to refute 
  KS's allegation, that the result is an impossibility proof. 
 }   
\begin{quote}
  This  word [`measurement'] very strongly suggests the ascertaining of some
pre-existing property of some thing, any instrument involved playing a purely
passive role.  Quantum experiments are just not like that, as we learned
especially from Bohr.  The results have to be regarded as the joint product
of `system' and `apparatus', the complete experimental set-up. \dots\ I am
convinced that the word `measurement' has now been so abused that the field
would be significantly advanced by banning its use altogether, in favour for
example of the word `experiment'. 
\end{quote}
In other words:  the Bell-KS theorem shows quantum mechanics to be so
extremely inconsistent with the ordinary idea of a measurement that it
would be better not to use the word ``measurement'' at all.  So Bell can
hardly be accused of understating the theorem's significance (at least in
this passage).

In Sections~\ref{sec:intro}--\ref{sec:Cabello} we showed that  Bell's
most important point---the point that measurements do not ascertain
pre-existing properties---remains valid.  We now need to examine  the
rest of what he says in the light of PMKC's discoveries.

When Bell wrote the above passage he had in mind the way that 
spin measurements work in the de Broglie-Bohm theory 
(Bell~\cite{Bell3} pp.~35 and~163; Dewdney \emph{et
al}~\cite{DewdneyA}; Holland~\cite{Holland}; Bohm and Hiley~\cite{Bohm2}).  In
that theory the apparatus interacts with the system, so as to manufacture a
value which did  not previously exist.  The unobservability of the pre-existing
values is a consequence of this.  So it appears to Bell that there is a
deep connection between the unobservability of the pre-existing values and
Bohr's point, concerning the importance of the complete experimental set-up.

This part of Bell's analysis clearly  does need revision.    In the PMKC models
the apparatus does passively reveal a value which was already there, as in
classical physics.  It is true that the experimenter does not acquire any
\emph{knowledge} thereby.  
So Bell's point, that one cannot \emph{ascertain} the
pre-existing values, still stands.    However, this is not because the
apparatus actively manufactures  completely different values.  Instead, it is
because the valuation is ``pathologically'' discontinuous.   

One might be tempted to conclude that Bohr and Bell were just wrong about 
 the importance of the complete experimental set-up.  However, that would
be incorrect.  The indivisibility of the system-apparatus complex, on which
Bohr so strongly insists, is still a feature of the PMKC models.  For
instance, we showed in Appleby~\cite{me1} that, in the case of a system
comprising three spin-$1/2$ particles, the very existence of a
property typically depends on the complete experimental 
set-up\footnote{
  It seems that a similar phenomenon occurs in Palmer's models~\cite{Palmer}. 
  Palmer's models are of some independent interest, 
  and should be examined by anyone concerned with these questions.
}
(also see the
discussion of sequential measurements on a single spin-$1$ particle in
Appleby~\cite{me2}).

So the PMKC models do not invalidate either of Bell's two main points
concerning the significance of the Bell-KS theorem. They do, however, have a
major impact on the logic of Bell's analysis.  As Bell sees it his
epistemological proposition (concerning our inability to know the
pre-existing properties) is a direct consequence of his Bohrian proposition
(concerning the importance of the complete experimental set-up).  It now
appears that he is misled by what turn out to be merely  accidental features
of the de Broglie-Bohm theory.  In the general case the two propositions are
independent of one another.

Of the two of them, it appears to us that the epistemological proposition is
the more important---which is why we said, at the beginning of this
section, that the implications of the Bell-KS theorem are
\emph{primarily} epistemological.  But let us now qualify that by saying
that the Bohrian proposition is by no means unimportant. In
particular, it is  connected with the non-locality of the MKC
models~\cite{me1}.

Finally, the PMKC models show that the Bell-KS theorem
(in its epistemological aspect) only applies to finite precision
measurements.  MKC misconstrue this implication  of their models: 
for they say that finite precision \emph{nullifies} the Bell-KS theorem.  In
fact, finite precision \emph{saves} the theorem.  It is infinite precision
measurements that would nullify the theorem, if we could perform them
(nullify its epistemological implications, that is---there would
still be the Bohrian aspect).

This is a very interesting result, which certainly puts quantum mechanics in
a different light.  It means that it is only the finite precision of real
laboratory instruments which prevents us giving a
\emph{non}-hidden  variables interpretation of quantum
mechanics (what Bell   calls an exposed variables
interpretation---see below).  However, one should not attach too
much significance to this point.  Even if  infinite precision instruments
existed our finite precision brains would be incapable of assimilating the
information they provided.  Moreover, even a quite modest number of
significant figures would take us below the Planck length---in which case it
would very likely be quantum mechanics itself that got nullified (and the PMKC
models along with it).

In Sections~\ref{sec:intro}--\ref{sec:Cabello} we defended Bell against MKC. 
The discussion in this section redresses the balance.  PMKC may not have
actually nullified Bell's argument.  But they have certainly exposed
some  serious defects.

For the sake of completeness we now present two further criticisms
of Bell, which are only indirectly motivated by PMKC's arguments.  We noted
earlier  that Bell is sympathetic to the de Broglie-Bohm theory.  This leads
him to underplay the significance of his contextuality theorem.   Much of the
confusion which has afflicted this subject is attributable to that.

Saying that the pre-existing values cannot all be ascertained by measurement
amounts to saying that some of those values must be hidden. So it may appear
that  the main implication of the Bell-KS theorem could be stated as follows: 
the theorem shows that there is no \emph{non}-hidden, or \emph{exposed}
variables interpretation of quantum mechanics.  Bell, however, is extremely
reluctant to admit that proposition:  for, as he says 
(Bell~\cite{Bell3}, p.92,
footnote~24, his italics),
\begin{quote}
  Pragmatically minded people can well ask \emph{why bother about hidden
entities that have no effect on anything?} 
\end{quote}
It should be noted that Bell is not just worried about what
pragmatically minded people might ask.  He feels the force of this objection
himself. In other words, he fears that his contextuality theorem is, if not
exactly a no-go theorem, at any rate something bordering on that.  So he tries
to find a way  of avoiding that conclusion.

In effect, Bell himself tries to nullify his own
theorem.
  His strategy  is   to
amputate
 the spin observables on which the proof 
is based\footnote{
  There are some similarities between this and MKC's approach.
  Bell and MKC both attempt to circumvent the theorem by 
  only assigning values to a restricted class of observables---a dense
  subset of $S^2$ in the case of MKC, positions in the case of Bell.
}.  
He thereby 
arrives at a kind of ``stripped-down'' version of the de Broglie-Bohm
theory, in which the only beables are the wave-function itself and the
particle  positions (Bell~\cite{Bell3}, pp.10, 34--5, 127--33, 160--3).  In
this picture ``the particle does not `spin', although the experimental
phenomena associated with spin are reproduced'' (\emph{ibid}, p.35).

Bell is under the  impression that, in the de Broglie-Bohm theory, measurements
of  position are always non-contextual. So  he thinks that a particle's true
position is empirically observable.  He considers that a particle has no other
intrinsic properties, apart from its position.  So it appears to him that
\emph{every} property is empirically observable.  He consequently thinks it
``absurd'' to describe the de Broglie-Bohm theory as a hidden variables theory
(Bell~\cite{Bell3}, p.201; also see
\emph{ibid}, pp.92, 128, 162--3).  He suggests that the term ``exposed
variables'' would be more  appropriate (\emph{ibid}, p.128). It is an opinion
that is  widely shared in the Bohmian community (see, for example, Bohm and
Hiley~\cite{Bohm2}, p.2 and Holland~\cite{Holland}, pp.106-7).

The first objection to this argument is that it is not in fact true that
measurements of de Broglie-Bohm position are always non-contextual.
 This is shown by  the discovery
(shortly after Bell's death) of Englert
\emph{et al}'s  ``surreal'' de Broglie-Bohm
trajectories~\cite{Englert,Englert2,surreal,DewdneyB}. 
In this phenomenon a particle's trajectory is recorded by an array of
detectors. If the detectors are only read after a non-zero time interval, then
it can happen that the particle is recorded as having been in one place when it
was in fact somewhere entirely different.   As Dewdney
\emph{et al}~\cite{DewdneyB} note, this is  ``yet another illustration of the
contextuality of measurements''.

The second objection is, to our mind, even more
telling.   
In Bell's ``stripped-down'' version of the de Broglie-Bohm theory objective
reality is ascribed to the entire
trajectory $x(t)$.  So Bell is wrong to think that the instantaneous position
$x$ is a particle's only intrinsic property.  The time derivative $dx/dt$ must
also be considered an intrinsic property.  And de Broglie-Bohm velocities are
\emph{generically} hidden. 
 Except in special cases they are completely different
from the empirical velocities, found by
measurement~\cite{Holland,Bohm,Bohm2,me5}.

Bell 
says that it is in  the ``\,`hidden'(!) variables'' that ``one finds an image
of the visible world'' (Bell~\cite{Bell3}, p.201).   This is perfectly true,
if by ``visible world'' is meant the macroscopic bodies of our ordinary
experience.   In the classical limit the variables are no longer
hidden. For instance, if one observes a bus as it journeys intermittently
down a London street, then it is the de Broglie-Bohm trajectory that registers
in one's brain.   However, that is a consequence of  the interaction between
the bus and its thermal environment (Bohm and Hiley~\cite{Bohm2}, chapter~8 and
Appleby~\cite{me4}).  Undecohered de Broglie-Bohm velocities tend to be
strikingly, and even grotesquely at variance with anything that is actually
observed.  For instance, the electron in an $ns$ state of a Hydrogen atom is,
according to the de Broglie-Bohm theory, always at rest.

For these reasons it appears to us that Bell's attempt to circumvent his
contextuality theorem is unsuccessful.  Of course, we have only considered
the de Broglie-Bohm theory.  One cannot, without more work, completely exclude
the possibility that there exists some other theory in which  positions
\emph{and} velocities are both empirically ascertainable.  But it seems
unlikely (see, for example, Clifton's~\cite{CliftonB} discussion of
KS~obstructions in the Weyl algebra).

So the Bell-KS theorem does  establish that no exposed variables
interpretation of quantum mechanics is possible.  At any rate, it 
strongly suggests that that is the case.

The point is non-trivial.  The
non-existence of an exposed variables interpretation of quantum mechanics is
often regarded as obvious.  However, the kind of semi-intuitive reasoning on
which that opinion is  based is not a  substitute for formal argument,
starting from the fundamental principles of quantum mechanics.  In any case a
little reflection suffices to show that the point is, in fact, very far
from obvious.  It certainly did not seem obvious to Bell.    Also, if the
point really were as obvious as is often supposed, then PMKC's
discovery, that it critically depends on the impossibility of performing
infinite precision measurements, would not have come as such a surprise.

This  brings us to our final criticism of Bell. 
 Kochen and Specker  see
the Bell-KS theorem as an impossibility proof.  That, of course, is wrong: 
Bohm commits no actual fallacy.  However, it appears to us
that Bell goes much too far in the opposite direction.
 Bell describes non-locality as a ``real problem'' (Bell~\cite{Bell3},
p.172).  By contrast, he sees contextuality as no kind of problem at
all.  In fact, he appears to regard the idea, that it should be seen as a
problem, as ``silly'' (see Mermin~\cite{MerminA}).  We will argue  that
this dismissive response is just as inappropriate as Kochen and Specker's
overly assertive one.

Bell takes this relaxed attitude because he thinks that contextuality only
limits our ability to ascertain pre-existing spins. Once one
appreciates that it also limits our ability to ascertain
pre-existing velocities and  (in certain circumstances) pre-existing
positions, then it becomes clear that contextuality does represent a  serious
problem.

The problem with a contextual theory just is the fact that the variables
are hidden.  This means that a contextual theory is, in a certain sense,
metaphysical.  Bell himself makes the point very clearly when he asks why we
should ``bother about hidden entities that have no effect on anything''
(Bell~§\cite{Bell3}, p.92).  Englert
\emph{et al}~\cite{Englert2} make the same point when they say (in connection
with the ``surreal'' de Broglie-Bohm trajectories mentioned above) ``if the
[de Broglie-Bohm] trajectories
\dots have no relation to the phenomena, in particular to the detected path
of the particle, then their reality remains metaphysical, just like the
reality of the ether of Maxwellian electrodynamics''.

Of course, a quantity does not need to be directly observable in order to be
physically relevant.   In conventional quantum mechanics the state vector of
an individual system is not directly observable.  Nevertheless, it plays an
essential role in the theory:  without it quantum mechanics could not
function as a predictive physical theory.  However, the pre-existing values
posited by a  hidden variables theory are not like that. 
There does not seem to be anything that can be calculated using such
values that cannot be calculated equally well without them.  They 
seem gratuitous.   That
is what is meant by calling them metaphysical.

We should acknowledge there are some points to be made on the other side.   It
is probably fair to say that Bell, in spite of what he explicitly says, is not
unaware of the considerations just adduced.  However, they are, for him,
outweighed by his
 aversion to the ``vagueness'' and ``subjectivity'' of the then
orthodox Copenhagen interpretation (Bell~\cite{Bell3}, p.160).  Bell feels
that the de Broglie-Bohm theory, though metaphysical, is at least
\emph{clear}.   One need not be a committed Bohmian to acknowledge the force
of that argument. Moreover, the fact that the hidden variables idea has, until
now,  proved to be   devoid of predictive power does not necessarily mean
that it will always remain so (see, for example, 
Valentini~\cite{ValA,ValB}, Farragi and Matone~\cite{Far} and  't
Hooft~\cite{tHooft}).  

Let us also note that the Bell-KS theorem only shows that one cannot,
for each observable $\hat{A}$,  identify the pre-existing value of
$\hat{A}$ with the outcome of a  finite precision quantum
measurement in the approximate direction of
$\hat{A}$.  It does not logically exclude the possibility  that one might
find out the value by some more sophisticated means.  

But, these qualifications aside, contextuality is clearly a problem,
in just  the same sense that Bell considers  non-locality to be a ``real
problem'' (Bell~\cite{Bell3}, p.172).     Furthermore, it is a problem  the
PMKC models do nothing to obviate.  It is  true that the PMKC models are not
contextual in the same way as more conventional theories, such as the  de
Broglie-Bohm theory.  But the essential difficulty remains.  The postulated
beables are still metaphysical. 

Indeed, it appears to us that it is the Bell-KS theorem which encapsulates
the really fundamental problem.  Non-locality simply provides a particularly
graphic illustration of this more basic  point, that the postulated level of
objective reality is systematically concealed from view.

Suppose, \emph{per impossibile}, that we could perform infinite precision
measurements.  Then it can be seen from the results proved in
Appleby~\cite{me1} that the MKC models would violate \emph{signal} 
locality\footnote{
   \emph{c.f.}\ Valentini's~\cite{ValA,ValB} speculation, that violations of
  signal
  locality might be observable in a large class of other hidden variables
  theories.
}.  
In that case non-locality would no longer be a conceptual 
problem.  It
would be an empirical 
prediction. 
 The prediction might be confirmed
(implying that relativity is wrong) or disconfirmed (implying that the MKC
models are wrong). Either way, we would not be involved in a cosmic
conspiracy.

But as it is we believe  non-locality  to be unobservable.
A theory which is profoundly non-local at the level of
the underlying beables somehow contrives to be completely local at the
level of the observable phenomena. This is certainly objectionable.  However,
it is only one illustration---albeit a very striking illustration---of the
more general point, that the postulated beables are highly metaphysical.

\section{Conclusion}
\label{sec:conclusion}
PMKC have made a most important contribution to this subject.  However, it is
not important for the reason  MKC think.  The PMKC models do not nullify
the Bell-KS theorem.  Instead, they give us a deeper and more accurate insight
into what the theorem is really telling us.

We have argued that the Bell-KS theorem has a  primarily epistemological
significance.  It concerns  the knowledge 
we  acquire by measurement.  So what one needs to ask is not:  ``how
much of
$S^2$ can be coloured
\emph{at all}?''  But rather:  ``how much of $S^2$ can be coloured in such a
way that the colours are  \emph{empirically knowable}?'' Once that is
understood it can be seen that there is no question of the theorem being
nullified.

It can also be seen that the theorem encapsulates the essential distinction
between quantum and classical.  Quantum mechanics does not (as was once
thought) require us to abandon the classical picture, of particles having
sharply-defined, fully objective properties.  However, it seems that we do
have to abandon the assumption, that the properties are empirically
knowable.  We can, if we like, retain the belief:  but only at the price of
making it metaphysical.  To Bell's ``pragmatically minded people'' a belief
of that kind  seems empty.

To some extent these points were already recognized by Bell and by others. 
However, Bell's account, as we have seen, is vitiated by a number of 
misconceptions.  PMKC's achievement is to devise models in which the essential
meaning of the theorem emerges in a particularly pure form.   This greatly
clarifies the issue.

Finally, let us note that the point, that the  distinction between classical
and quantum is  partly epistemological in character, acccords
with the current interest in quantum 
information\footnote{
  In this connection let us note that MKC suggest that their
  models may have implications for quantum computation.
  However, they are misled by their belief that ``once the assumption
  of infinite precision is relaxed'' non-relativistic quantum mechanics
 can be simulated classically (Clifton and Kent~\cite{Kent2}, p.2103).  
  As we have seen (in Section~\ref{sec:Significance}) their models actually
  suggest the exact opposite:  namely,
 that it is only if infinite precision measurements \emph{were} possible
 that quantum mechanics might be simulated classically (in which case it would
 also
  violate
  \emph{signal}  locality).
  As Meyer himself remarks (citing Sch\"{o}nhage~\cite{Schonhage} and
 Freedman~\cite{Freedman}) it would not be
  surprising if  a quantum computer performed no better than an 
  \emph{infinite} precision classical machine.
}.  
In particular, it accords with Fuchs's idea~\cite{Fuchs,Caves1}, that quantum
mechanics can partly (and perhaps even mostly) be seen as a ``law of thought''.

In this paper we have tried to avoid taking sides in the
interpretational dispute.   But, now that we have reached the end, let us say
that we share the pragmatically minded person's distaste for metaphysical
theories.  On the other hand, we also share Bell's distaste  for the
vagueness and subjectivity of the Copenhagen interpretation.  It appears to
us that what Bell says on that score is amply justified.  We therefore find
ourselves impaled on the horns of a very unpleasant dilemma.

We do not profess to know how the dilemma can be resolved.  But one
possibility would be to improve the Copenhagen interpretation to the point
where it was no longer vague, and no longer subjective.  Or, at any rate, not
so offensively vague, and not so offensively subjective.  Fuchs's
``law of thought'' idea strikes us as very promising in that respect. 
However, it would take a great deal of work before that promise could be
fulfilled.

\subsubsection*{Acknowledgements}  The author is grateful to H.~Brown,
J.~Butterfield, A.~Cabello,
R.~Clifton, C.A.~Fuchs, A.~Kent, J.\AA.~Larrson, N.D.~Mermin,
T.N.~Palmer, A.~Peres,
I.~Pitowsky and K.~Svozil for useful discussions.

\appendix

\section{Proof of Lemma~\ref{lem:KSStrongerB}}
\label{sec:Lem2Proof}
In this appendix we prove that $d_{\mathcal{B}} >0$, as
stated in Lemma~\ref{lem:KSStrongerB}.  We  conclude with a few remarks
concerning its actual magnitude.

Let $\{\mathbf{n}_1, \mathbf{n}_2, \dots , \mathbf{n}_{2 M}\}$ be a
KS-uncolourable 
set\footnote{
  In view of the way we defined  KS-colourable sets in
  Section~\ref{sec:Preliminaries} (also see the footnote in
  Section~\ref{sec:PseudoKS}) the proof of uncolourability
  must not make any \emph{independent} appeal to the requirement
  that linear combinations of vectors evaluating to $1$ should
  also evaluate to $1$.
 }
of unit vectors with the property
$\mathbf{n}_{M+i}=-\mathbf{n}_{i}$ for $i=1,\dots,M$.  Let $\theta_0$ be the
minimum angular separation of the vectors in this set:
\begin{equation}
\theta_0 = \min_{1\le i,j\le 2 M}\left(\cos^{-1}\left(\mathbf{n}_i \cdot
\mathbf{n}_j\right)
\right)
\end{equation}
For each $i$, surround $\mathbf{n}_i$ with a circular patch $E_i$
of radius
$\theta_0/2$:
\begin{equation}
E_i = \{\mathbf{m} \in S^2\colon \cos^{-1} \left(\mathbf{m} \cdot
\mathbf{n}_i\right) \le \theta_0/2\}
\end{equation}

 Let $B$ be a set $\in \mathcal{B}$, and let $B^{c}=S^2-B$ be its
complement. We may assume, without loss of generality, that $B$ (and
consequently $B^{\mathrm{c}}$) is invariant under the parity operation
(since, if $B$ is not invariant, we can  replace it with another
set $\in
\mathcal{B}$ which is invariant, and whose measure is the same or larger).

By construction the sets $E_i$ are non-overlapping with the possible
exception of a set of measure zero on their boundaries.  Consequently
\begin{equation}
\mu(B^c) \ge \sum_{i=1}^{2 M} \mu \left(E_i \cap B^c\right)
\label{eq:BcBoundA}
\end{equation}

We now define, for each $i$, a function $g_i\colon S^2\to E_i$ by 
\begin{equation}
g_i (\mathbf{m}) = e^{(\theta_0/2) \mathbf{m} \cdot \mathbf{L}} \mathbf{n}_i
\end{equation}
for each $\mathbf{m} \in S^2$.  Here  $L_1, L_2, L_3$ are the generators of
$SO(3)$.  Thus $g_i$ maps $\mathbf{m}$ onto the vector obtained by rotating
$\mathbf{n}_i$ through the  (fixed)
angle $\theta_0/2$ about the (variable) axis $\mathbf{m}$ .  It is easily
seen that, as $\mathbf{m}$ ranges over  $S^2$, the interior of
$E_i$ is covered twice, and the boundary once.  Consequently
\begin{equation}
\mu\left(E_i \cap B^c\right) =
\frac{1}{2} \int_{\tilde{B}_i^c} J_i (\mathbf{m}) \, d \mu
\label{eq:EBmeasTermsJi}
\end{equation}
where $J_i$ is the Jacobian of $g_i$ and $\tilde{B}_i^c=g_i^{-1} \left(E_i
\cap B^c\right)$.

Now let 
\begin{equation}
J(\mathbf{m}) = \min_{1\le i \le 2 M} \left( J_i (\mathbf{m})\right)
\end{equation}
Eqs.~(\ref{eq:BcBoundA}) and~(\ref{eq:EBmeasTermsJi}) then imply
\begin{equation}
\mu(B^c) \ge \frac{1}{2}\int_{\cup_{i=1}^{M}\tilde{B}_i^c} J(\mathbf{m}) \,
d\mu + \frac{1}{2}\int_{\cup_{i=M+1}^{2 M}\tilde{B}_{i}^c} J(\mathbf{m}) \,
d\mu
\end{equation} 

We now observe that, for each fixed value of $\mathbf{m}$, the set 
$\{g_1(\mathbf{m}), \dots , g_{2 M} (\mathbf{m})\}$, being obtained  by
rotating the KS-uncolourable set $\{\mathbf{n}_1, \dots , \mathbf{n}_{2
M}\}$, must itself be KS-uncolourable.  Since $B$ is KS-colourable this means
that, for each $\mathbf{m}$, there must exist some $1\le i \le M$ such that
$g_i(\mathbf{m}) $ and $g_{i+M} (\mathbf{m})$ both $\in B^c$.  Consequently 
$\cup_{i=1}^{M} \tilde{B}_i^c=\cup_{i=1}^{M} \tilde{B}_i^c=S^2$.  Hence
\begin{equation}
\mu(B^c) \ge \int_{S^2} J(\mathbf{m}) d \mu
\end{equation}

We deduce that
\begin{equation}
\mu (B) \le 1 - \int_{S^2} J(\mathbf{m}) \, d \mu
\end{equation}
for all $B \in \mathcal{B}$.  
Finally, we note that $J(\mathbf{m})$ is a continuous, non-negative function
which is not identically $0$.  This fact, together with 
Eq.~(\ref{eq:dBdef}), implies
\begin{equation}
d_{\mathcal{B}} \ge \int_{S^2} J (\mathbf{m}) \, d\mu >0 
\label{eq:dBBound}
\end{equation}
This proves Lemma~\ref{lem:KSStrongerB}.

Finally, let us briefly consider the size of this integral.  An exact
calculation, though straightforward, would be somewhat tedious.  We therefore 
confine ourselves to noting that it follows from the definition of $J$ that,
for all $i$,
\begin{equation}
\int_{S^2} J (\mathbf{m}) \, d\mu \le 
\int_{S^2} J_i (\mathbf{m}) \, d\mu
=2 \mu(E_i) = 2 \sin^2 \left(\frac{\theta_0}{4}\right)
\end{equation}
For the Conway-Kochen set~\cite{PeresBk,BubBk} one has
$\theta_0=18.4^{\mathrm{o}}$ (the angle between the directions $(0,1,2)$ and
$(0,2,2)$), implying that for this set
\begin{equation}
\int_{S^2} J (\mathbf{m}) d\mu < 0.013
\end{equation}
If the Conway-Kochen set  maximizes the integral, and if 
$d_{\mathcal{B}}$ is of the same order as the lower bound set by
Inequality~(\ref{eq:dBBound}), it would follow that $d_{\mathcal{B}}\lesssim
0.01$.  However, it would require further investigation to tell whether that
is actually the case.
\section{Proof of Lemma~\ref{lem:LebesgueVitali}}
\label{ap:LebesgueVitali}
The lemma is a consequence of the Lebesgue-Vitali theorem (see, for example,
Shilov and Gurevich~\cite{Shilov}, Chapter~10).

Let $\phi$  be an integrable function defined on a measure space
$(X,S,\mu)$ with $\sigma$-ring $S$ and countably additive measure $\mu$. 
Then it can be shown (Shilov and Gurevich~\cite{Shilov}, pp.~220--1)  that,
for almost all $x_0\in X$, the limit
\begin{equation}
  \lim_{\delta \to 0} \left( \frac{1}{\mu \bigl(V_{\delta}(x_0)\bigr)}
 \int_{V_{\delta}(x_0)} |\phi(x) - \phi(x_o)|\, d\mu
\right)
\label{eq:Vitali}
\end{equation} 
exists and $=0$, provided that for each $\delta > 0$, $V_{\delta} (x_0)$
is a Vitali set containing $x_0$ and having $\mu$-measure $< \delta$ (for
the definition of a Vitali set  see Shilov and
Gurevich~\cite{Shilov}, p.~209).

A straightforward modification of Banach's elegant argument to show that the
set of cubes is a Vitali system for $\mathbb{R}^n$ (Shilov and
Gurevich~\cite{Shilov}, pp.~216--8) establishes that the set of disks
$S(\mathbf{n},
\epsilon)$ is a Vitali system for $S^2$. 

The result is now immediate if we take  $\phi$ in Eq.~(\ref{eq:Vitali})
to be the indicator function of $D$ (\emph{i.e.}\ the function which is 
$1$ on $D$ and $0$ on its complement).

\end{document}